\documentclass[aps,prb,twocolumn,superscriptaddress,floatfix,10pt,nofootinbib]{revtex4-2}
\usepackage[utf8]{inputenc}
\usepackage{natbib}
\usepackage{graphicx}
\usepackage{latexsym}
\usepackage{amssymb}
\usepackage{amsmath}
\usepackage{amsfonts}
\usepackage{amsthm}
\usepackage{bm}
\usepackage{floatrow}
\usepackage{subfig}
\usepackage{bbm}
\usepackage{enumitem}
\usepackage{hyperref}
\usepackage{lipsum}
\usepackage{braket}
\usepackage{tikz}
\usepackage{pgfplots}
\usepackage{ifthen}
\usepackage{caption}
\usepackage{ragged2e}
\usepackage{multirow}
\usepackage[export]{adjustbox}
\usetikzlibrary{arrows,decorations.pathreplacing,decorations.markings,arrows.meta,patterns,3d}
\usepackage[normalem]{ulem} % for strikeout \sout https://tex.stackexchange.com/a/23712
\usepackage{cancel}

\theoremstyle{definition}

\pgfplotsset{compat=1.8}

\begin{document}

\title{String operators for Cheshire strings in topological phases}
\date{\today}

\author{Nathanan Tantivasadakarn}
\affiliation{Walter Burke Institute for Theoretical Physics and Department of Physics, \mbox{California Institute of Technology, Pasadena, CA, 91125, USA}}
\author{Xie Chen}
\affiliation{Walter Burke Institute for Theoretical Physics and Department of Physics, \mbox{California Institute of Technology, Pasadena, CA, 91125, USA}}
\affiliation{Institute for Quantum Information and Matter, \mbox{California Institute of Technology, Pasadena, CA, 91125, USA}}

\begin{abstract} 
Elementary point charge excitations in 3+1D topological phases can condense along a line and form a descendant excitation called the Cheshire string. Unlike the elementary flux loop excitations in the system, Cheshire strings do not have to appear as the boundary of a 2d disc and can exist on open line segments. On the other hand, Cheshire strings are different from trivial excitations that can be created with local unitaries in 0d and finite depth quantum circuits in 1d and higher. In this paper, we show that to create a Cheshire string, one needs a linear depth circuit that acts sequentially along the length of the string. Once a Cheshire string is created, its deformation, movement and fusion can be realized by finite depths circuits. This circuit depth requirement applies to all nontrivial descendant excitations including symmetry-protected topological chains and the Majorana chain.
\end{abstract}

\maketitle

``Cheshire strings'' describe line-like excitations in topological states which contain a mysterious hidden charge along the length of the string\cite{Alford1990, Preskill1990}. The charge content cannot be pinpointed to one or a few local regions on the string and therefore costs no extra energy compared to the zero charge state. The degeneracy is a result of the spontaneous symmetry breaking, or more precisely `Higgsing', of (part of) the gauge symmetry in the topological phase and the condensation of the corresponding gauge charges along the string. Since the gauge charge is condensed, adding more does not change the energy of the condensate. 

Cheshire strings have been included as an essential part in the formulation of a complete mathematical description of $3+1$D topological orders in terms of higher categories\cite{Gaiotto2019arxiv, Kong2020, Else2017, Kong2014arxiv, Douglas2018arxiv, Lan2018, Kong2020center, Johnson-Freyd2022, Kong2022, Kong2020defects, Kong2022arxiv_liquids}. In Ref.~\onlinecite{Else2017}, the example of $3+1$D Toric Code was discussed explicitly where fundamental objects in the higher category include $1$d objects like the Cheshire string and magnetic gauge flux loop and the $0$d domain walls between them which includes the electric gauge charge excitation. From a physics perspective, it is not immediately clear why a consistent theory of the $3+1$D $\mathbb Z_2$ topological order needs to involve the Cheshire string. After all, the elementary excitations -- the gauge charge and gauge flux -- capture all the fractional statistics we see in the model and their corresponding string and membrane operators capture the full logical operations in the degenerate ground space. The Cheshire string is made of gauge charges and in this sense a `descendant' excitation\cite{Kong2014arxiv}. Are they really necessary in the description?

Some observations point to the intrinsic nature of Cheshire strings in $3+1$D topological phases. First of all, they do exist as a $1d$ excitations in the model and can hence appear next to magnetic flux loops and change their feature. In non-abelian gauge theories, it has long been known that magnetic flux loops are intrinsically ``Cheshire" as the flux loop breaks the non-abelian gauge symmetry down to a subgroups and therefore gauge charges that transform nontrivially under the broken part of the gauge group have to condense along the loop \cite{Alford1990, Preskill1990}. In abelian gauge theories with nontrivial three-loop braiding\cite{Wang2014,Jian2014,Bi2014,Jiang2014,Wang2015,Wang2015a,Chen2016,Cheng18}, some flux loops are intrinsically ``Cheshire'' as well. Such flux loops can be thought of as the boundary between different $2+1$D gauge theories and can only be gapped if the gauge charge is condensed along the loop\cite{Else2017}. 

The following Q$\&$A may help clarify the situation.

\begin{itemize}

\item Do we need to include Cheshire \textit{points} in our description of topological order as well? 

We already have. A Cheshire point can be obtained by shrinking the Cheshire string to a point and is a direct sum of all charge states. In the $\mathbb Z_2$ case, it is $1 \oplus e$. When the condensate is $0$d, the degeneracy between the two charge states will generically split and we end up with either $1$ or $e$. 

\item Why did we not include Cheshire strings in the description of $2+1$D topological order?

$2+1$D topological orders are characterized in terms of braided fusion categories whose fundamental objects are point excitations -- the anyons. Cheshire strings appear at one higher dimension and hence do not mix with the anyons. However, they do play an important role as defects and boundaries of the topological states and are described as  (bi)module categories over the bulk fusion category, as explained in Ref.~\onlinecite{Kitaev2012}.

\item Apart from Cheshire strings, are there other types of descendant excitations? 

Yes. Cheshire strings correspond to the $1+1$D symmetry breaking phase of the gauge charge. There are also invertible phases such as symmetry-protected topological phases in $1+1$D. For example, when the gauge charge is a fermion, there is also the Majorana chain. Cheshire strings are a non-invertible excitations while the latter two correspond to invertible descendant excitations. (Refs.~\onlinecite{Yoshida2015,Yoshida17,roumpedakis2022higher,Kong2022arxiv,Barkeshli23} construct examples of such excitations.) Going to higher dimensions, we would also need to consider membrane-like descendant excitations of point and loop excitations.

\item What is the benefit of including Cheshire string in the description of topological orders?

One benefit is that it makes the correspondence between bulk and boundary more natural (at least in a mathematical sense as shown in Refs.~\onlinecite{Kitaev2012,Kong2020center, Kong2017, Kong2015arxiv}).

\item The elementary excitations are generated with unitary string and membrane operators. Does this also apply to Cheshire string?

Yes, this is what we are going to show in this paper. Previous discussions have mostly focused on generating Cheshire string by changing the Hamiltonian of the topological state or using projection operations\cite{Else2017}. We want to emphasize that Cheshire strings can also be generated using unitary circuits and the way it is generated using unitary circuits makes clear its nontrivialness as a descendant excitation. Magnetic fluxes are nontrivial loop excitations because they can only be generated as the boundary of a unitary membrane operator. Because of that, they have to form closed loops. Cheshire strings, on the other hand, do not have to appear on the boundary of a membrane and can exist on open string segments (i.e. they admit a topological boundary). However, if we would like to generate a Cheshire string with a unitary circuit along the string, the circuit depth has to grow linearly with the length of the string\footnote{An argument is the following: suppose there exists a finite depth circuit to create a Cheshire string, we may use it to create  two Cheshire strings of extensive length and extensive separation in the system size. This configuration now has an additional ground state degeneracy. The degeneracy can be labeled by the amount of charge on each Cheshire string, which can only be detected non-locally by moving a flux around one of the Cheshire strings. Moreover, changing the amount of charge on each Cheshire string requires an operator of extensive support which is the string operator of the charge. Since this degeneracy is robust, it is a contradiction.}. The linear depth of the circuit distinguishes Cheshire strings from trivial line-like excitations that can be generated with finite depth circuits. In particular, as we show below, the linear depth circuit has a sequential structure. That is, each layer contains only one local gate set acting on a local region in the string. The gate set moves from one end point of the string to the other and acts on the string in sequential way. This is hence an example of a Sequential Quantum Circuit\cite{Schon2005,Schon2007,Banuls2008,Wei2022,Chen2023}. Once a Cheshire string has been created, we show that it can be deformed, moved, and fused using finite depth circuits. Therefore, the equivalence relation between Cheshire excitations can be established with finite depth circuits, similar to the case of elementary point and loop excitations\footnote{Equivalence relations between point excitations are given by local unitary transformations.}. 

\end{itemize}

The paper is structured as follows: in Section~\ref{sec:2Dgenerate}, we show how to generate a Cheshire string in $2+1$D Toric Code using a sequential linear depth circuit; in Section~\ref{sec:2Dfuse}, we show how to deform, move and fuse Cheshire strings in $2+1$D Toric Code; in Section~\ref{sec:3D}, we show how similar circuits work in $3+1$D. In fact, in the $2+1$D Toric Code, there are five types of nontrivial defects\cite{Kong2022arxiv, roumpedakis2022higher}. Their creation and fusion can be achieved similarly with sequential linear depth circuits / finite depth circuits. We demonstrate this in Appendix~\ref{sec:defects}. In the summary section (section~\ref{sec:summary}), we discuss the relation between different types of excitations in topological phases and the unitary operation that generates them.

\begin{figure}[b]
    \centering
    \includegraphics[scale=0.45]{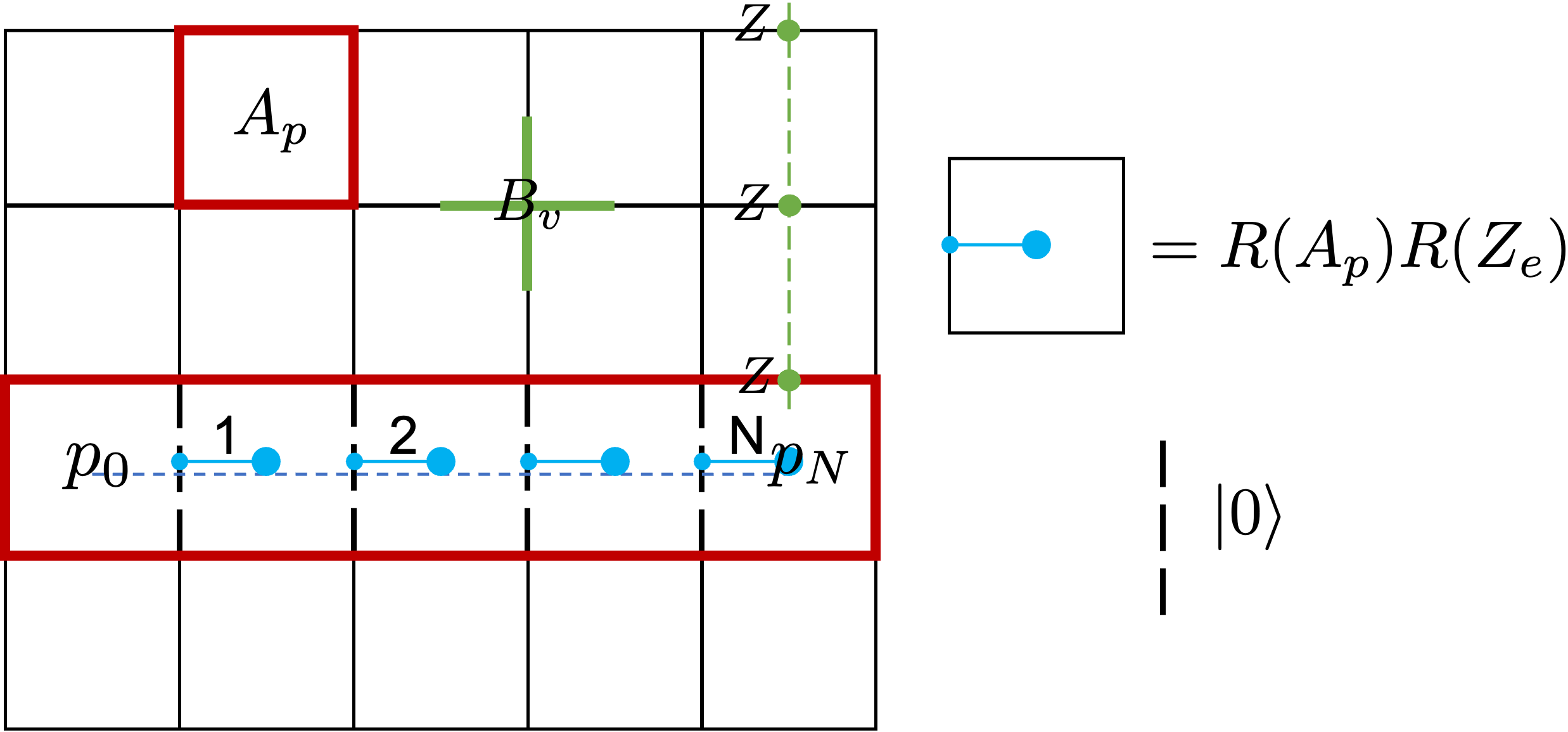}
    \caption{Generation of an $e$-Cheshire string in $2+1$D Toric Code with a sequential circuit. $A_p$ and $B_v$ are Hamiltonian terms of the $2+1$D Toric Code. A Cheshire string (on the dual lattice) from $p_0$ to $p_N$ is generated by applying a sequence of gate sets represented by the blue dot pairs. The dashed black edges are mapped to the product state $|0\rangle$ forming the condensate, while the total charge of the condensate measured by $\prod X$ around the red loop is conserved. Adding an $e$ charge using the string of $Z$ along the dashed green line maps between the two degenerate states of the condensate.}
    \label{fig:2DSe}
\end{figure}

\section{Generation of Cheshire string in 2+1D}
\label{sec:2Dgenerate}

Let us first consider the generation of a Cheshire string in the $2+1$D Toric Code. Consider the Toric Code on two dimensional lattice defined with the Hamiltonian
\begin{equation}
\begin{array} {lll}
H & = & -\sum_p A_p - \sum_v B_v \\
& = &
-\sum_p \prod_{e\in p} X_e - \sum_v \prod_{v \in e} Z_e
\end{array}
\end{equation}
Let's call excitation of the $A_p$ terms the gauge charge excitation labeled by $e$ and the excitation of the $B_v$ terms the gauge flux excitation labeled by $m$. Applying $Z_e$ on one edge creates two gauge charge excitations on the neighboring plaquettes. Having a charge condensate corresponds to enforcing $-Z_e$ as the Hamiltonian term so that the ground state remains invariant under the pair creation or hopping of gauge charges between the neighboring plaquettes. If such a term is enforced on a string of edges on the dual lattice (dotted blue line in Fig.~\ref{fig:2DSe}), we get a Cheshire string for the gauge charge $e$.

While the Cheshire string can be generated with projection operators $\frac{1}{2}\left(1+Z_e\right)$ acting on the edges along the string all at the same time, generating it with unitary operations takes a number of steps that scales linearly with the length of the string. Consider creating a Cheshire string on the dual lattice, which starts at a plaquette $p_0$ and ends at plaquette $p_N$, adjacent plaquettes in the sequence $p_i$ and $p_{i+1}$ share an edge $e_{i,i+1}$. The sequential circuit to create the Cheshire charge is given by 
\begin{align}
   U= \prod_{i=N}^{1} R( A_{p_i})R(Z_{e_{i-1,i}})
\end{align}
where we define $R(\mathcal O) \equiv  e^{-\frac{i\pi}{4} \mathcal O}$, which has the property that for Pauli operators $P$ and $Q$,
\begin{align}
    R(Q) P R(Q)^\dagger = \begin{cases}
    P ;& [P,Q]=0\\
    iPQ; & \{P,Q \}=0
    \end{cases}
\end{align}
Note that the $R$ gates do not commute with each other so the ordering is important. Our convention is that gates to the right are applied before gates to the left. Therefore, the sequence of gates are $R(Z_{e_{0,1}})$, $R(A_{p_1})$, $R(Z_{e_{1,2}})$, $R(A_{p_2})$, and so forth. Since $A_p$ and $Z_e$ commutes with the vertex term, the circuit leaves the vertex stabilizers invariant while mapping
\begin{align}
    A_{p_i} \to Z_{e_{i-1,i}},\ i = 1,...,N\\ \nonumber
    A_{p_0} \to \prod_{i=0}^N A_{p_i}\prod_{i=1}^N Z_{e_{i-1,i}}
\end{align}
Therefore, after the circuit, the edges along the dual string are in the condensate state $|0\rangle$ stabilized by $Z_e$. The total charge of the condensate measured by $\prod_{i=0}^N A_{p_i}$ remains invariant in the ground state, while the individual charges $A_{p_i}$ are no longer conserved. Adding an $e$ charge using the string of $Z$ operators along the dashed green line changes the total charge of the condensate, but does not affect the local terms in the condensate. Therefore, the two total charge states are degenerate on the Cheshire string.

Generating a Cheshire string -- a condensate for the gauge charge $e$ -- in the Toric Code corresponds to opening up a slit of vacuum state inside the topological bulk with a `smooth' boundary between the two. Similar operations (or the inverse) have been discussed in Refs.~\onlinecite{Dennis2002,Aguado2008,Liu2022}. In appendix~\ref{sec:defects}, we are going to discuss how a similar condensate of the gauge flux $m$ corresponds to a `rough' boundary between Toric Code and the vacuum. 

\section{Fusion of Cheshire string in 2+1D}
\label{sec:2Dfuse}

Once a Cheshire string is created, it can be deformed, moved, and fused using finite depth circuits. In other words, finite depth circuits establish the equivalence relation between Cheshire string excitations. 

\begin{figure}[ht]
    \centering
    \includegraphics[scale=0.43]{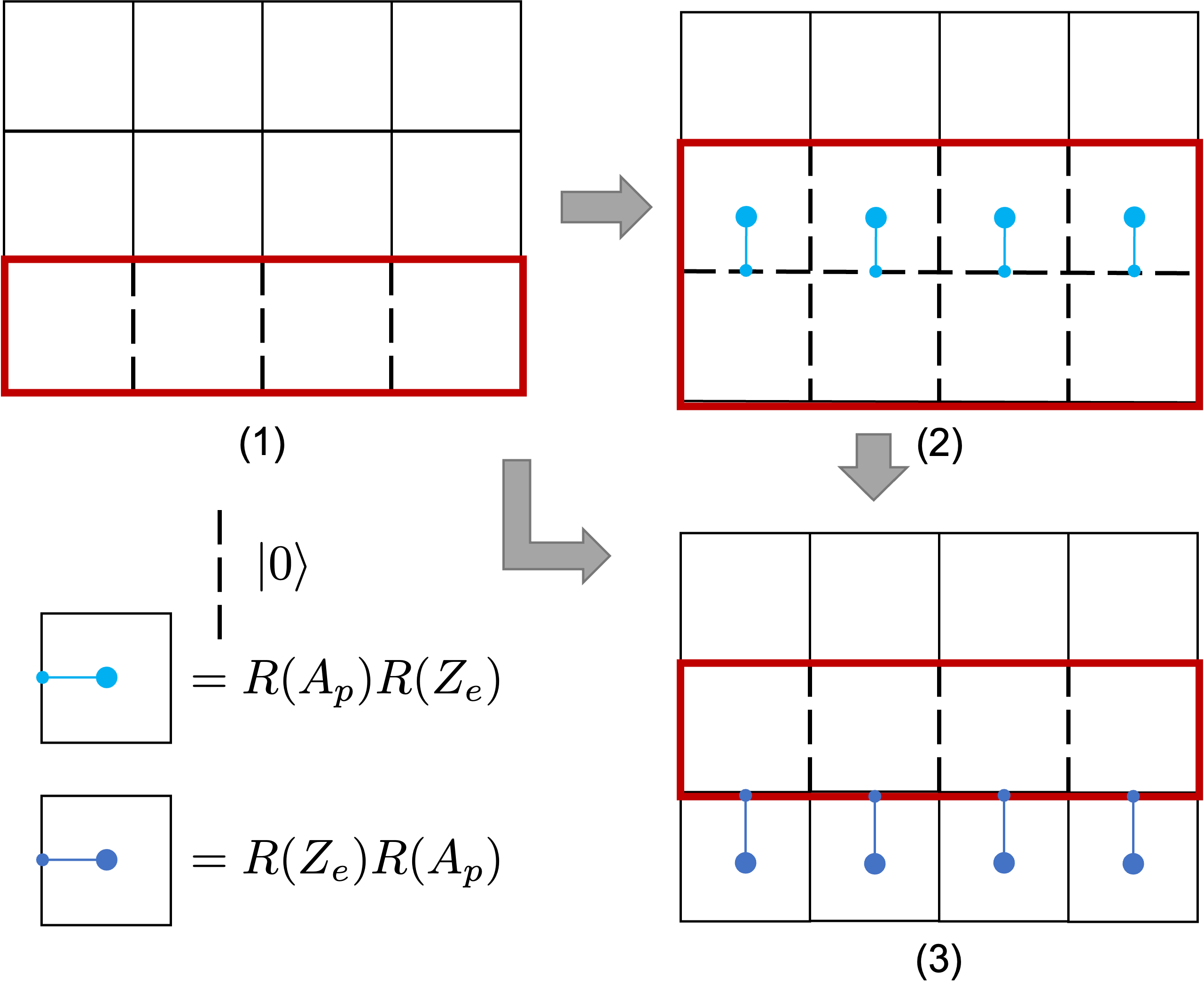}
    \caption{Deforming ((1) to (2) and (2) to (3)) and moving ((1) to (3)) a Cheshire string using finite depth circuit in $2+1$D. The gate sets in each diagram can be applied in parallel. The dashed black edges are in the $|0\rangle$ state of the condensate. Red loop of $\prod X$ measures the total charge in the condensate.}
    \label{fig:2DSe_move}
\end{figure}

As shown in Fig.~\ref{fig:2DSe_move}, if we start from a Cheshire string in the bottom row ((1) with dashed black edges in state $|0\rangle$), we can make it thicker (2) and then thinner (3) using finite depth circuits. The individual gate sets (the blue dot pairs) take the same (inverse) form as in Fig.~\ref{fig:2DSe}. The difference is that, now the gate sets are oriented in parallel, rather than connecting head to toe. It can be easily checked that parallel gate sets commute with each other and hence can be applied simultaneously. Going from (1) to (3) moves the Cheshire string perpendicular to its length by one step. The total charge of the condensate, measured by $\prod X$ along the red loop, is conserved in the whole process. 

\begin{figure}[ht]
    \centering
    \includegraphics[scale=0.45]{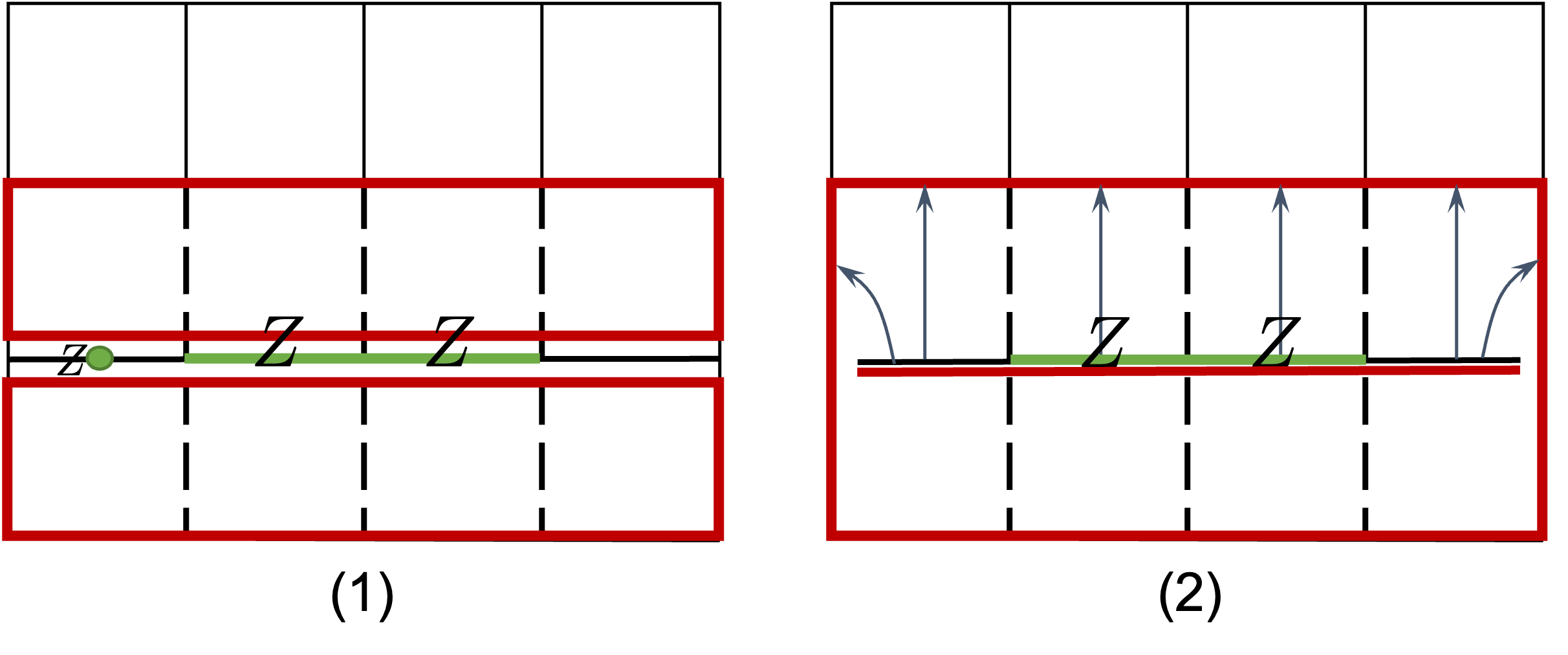}
    \caption{Fusion of two Cheshire strings with a finite depth circuit in $2+1$D. (1) Before fusion, $\prod X$ around each red rectangle measures the total charge on each string. Qubits on the domain wall are coupled with $ZZ$ terms (green bond). A single $Z$ on the domain wall (green dot) tunnels a charge between the two condensates. (2) Applying the controlled-NOT gates indicated by the arrows, the domain wall qubits are decoupled from the bulk. The $ZZ$ coupling remains, giving rise to two fold degeneracy.}
    \label{fig:2DSe_fuse}
\end{figure}

Two identical Cheshire strings $c$ defined on a segment fuse as\footnote{More generally, the coefficient of the fusion depends on the spacetime manifold of the string\cite{roumpedakis2022higher}. Explicitly, for a Cheshire string supported on a spatial manifold $M$, the coefficient is the partition function of a 1+1D $\mathbb Z_2$ gauge theory on $M \times S^1$. I.e. it is the degeneracy of a 1+1D $\mathbb Z_2$ Ferromagnet on $M$.}
\begin{equation}
 c \times c = 2c
\end{equation}
which can be realized with finite depth circuit as well. Using the circuits discussed above, we can always move the two strings with finite separation right next to each other as shown in Fig.~\ref{fig:2DSe_fuse}(1) with a finite depth circuit. 

Before fusing, $\prod X$ around the red rectangles measures the charge on each string. Along the domain wall separating the two strings, the Hamiltonian involves pairwise $ZZ$ terms as shown by the green bond in Fig.~\ref{fig:2DSe_fuse}(a). A single $Z$ on the domain wall tunnels a charge from one string to another. It spoils the conservation of charge on each string while preserving their sum. 

The fusion of the strings can be achieved with the finite depth circuit shown in Fig.~\ref{fig:2DSe_fuse}(b). Each arrow represents a controlled-NOT gate with the start of the arrow being the control and the end of the arrow being the target. All the gates commute and can be applied in one step. After the circuit, the domain wall separating the two strings become completely decoupled from topological bulk. The $ZZ$ couplings along the domain wall remain invariant, resulting in a two fold degeneracy between the $|00...0\rangle$ state and the $|11...1\rangle$ state. When the domain wall is in the $|00...0\rangle$ state, it merges naturally with the condensates on the two sides into a single condensate. When the domain wall is in the $|11...1\rangle$ state, a simple one step rotation would take it into the $|00...0\rangle$ state and the same conclusion holds. Therefore, two Cheshire strings merge into one Cheshire string with a pre-factor of 2. If one wants to put the fused string into a standard form (e.g. of width 1), this can be done with another finite depth circuit of the form in Fig.~\ref{fig:2DSe_move}(3).

\section{Cheshire string in 3+1D}
\label{sec:3D}

\begin{figure}[ht]
    \centering
    \includegraphics[scale=0.43]{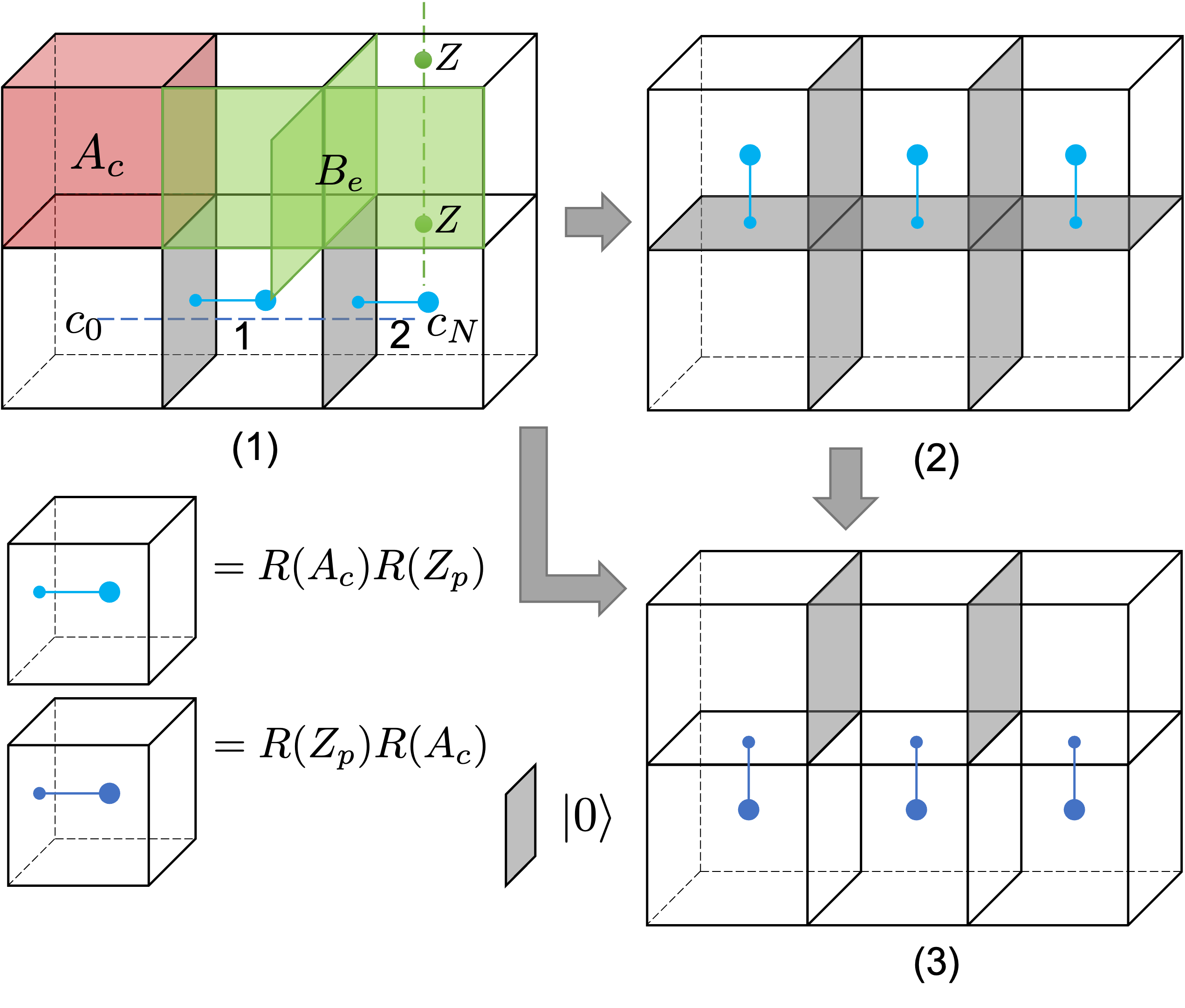}
    \caption{Creating, deforming and moving Cheshire string in $3+1$D Toric Code. $A_c$ and $B_e$ are Hamiltonian terms in the Toric Code. (1) Cheshire string along the dual string (blue dotted line) can be generated with a sequential circuit along the length of the string. Once the string is created, it can be deformed ((1) to (2) and (2) to (3)) and moved ((1) to (3)) with finite depth circuits.}
    \label{fig:3DSe}
\end{figure}

A similar construction holds in $3+1$D Toric Code as well. Consider the $3+1$D Toric Code on a cubic lattice with $Z_2$ qubits on the plaquettes. The Hamiltonian contains a charge term $A_c$ around each cube $c$ and a flux term $B_e$ around each edge $e$, as shown in Fig.~\ref{fig:3DSe}(1).
\begin{equation}
\begin{array} {lll}
H & = & -\sum_c A_c - \sum_v B_v \\
& = &
-\sum_c \prod_{p\in c} X_p - \sum_e \prod_{e \in p} Z_p
\end{array}
\end{equation}

Similar to the $2+1$D case, applying $Z_p$ on one plaquette creates two gauge charge excitations in the neighboring cubes. A Cheshire string corresponds to enforcing $-Z_p$ on the plaquettes along a dual string (the dotted blue line in Fig.~\ref{fig:3DSe}(1)). 

As shown in Fig.~\ref{fig:3DSe}(1), a Cheshire string from cube $c_0$ to $c_N$ can be created with a sequential circuit
\begin{equation}
U = \prod_{i=N}^1 R(A_{c_i})R(Z_{p_{i-1,i}})
\end{equation}
The circuit leaves the $B_e$ terms invariant while mapping
\begin{align}
    A_{c_i} \to Z_{p_{i-1,i}},\ i = 1,...,N\\ \nonumber
    A_{c_0} \to \prod_{i=0}^N A_{c_i}\prod_{i=1}^N Z_{p_{i-1,i}}
\end{align}
Similar to the $2+1$D case, the shaded plaquettes are mapped to the $|0\rangle$ state of the condensate while the total charge of the condensate measured by $\prod_{i=0}^N A_{c_i}$ remains invariant in the ground state. Adding an $e$ charge using the string of $Z$ operators along the dashed green line maps between the two degenerate states of the Cheshire string. By generating a Cheshire string in the bulk of the Toric Code state, we create the vacuum state in a line like region with a smooth gapped boundary to the topological bulk. Similar operations (or the inverse) have been discussed in Ref.~\onlinecite{Chen2022arxiv}.

\begin{figure}[t]
    \centering
    \includegraphics[scale=0.43]{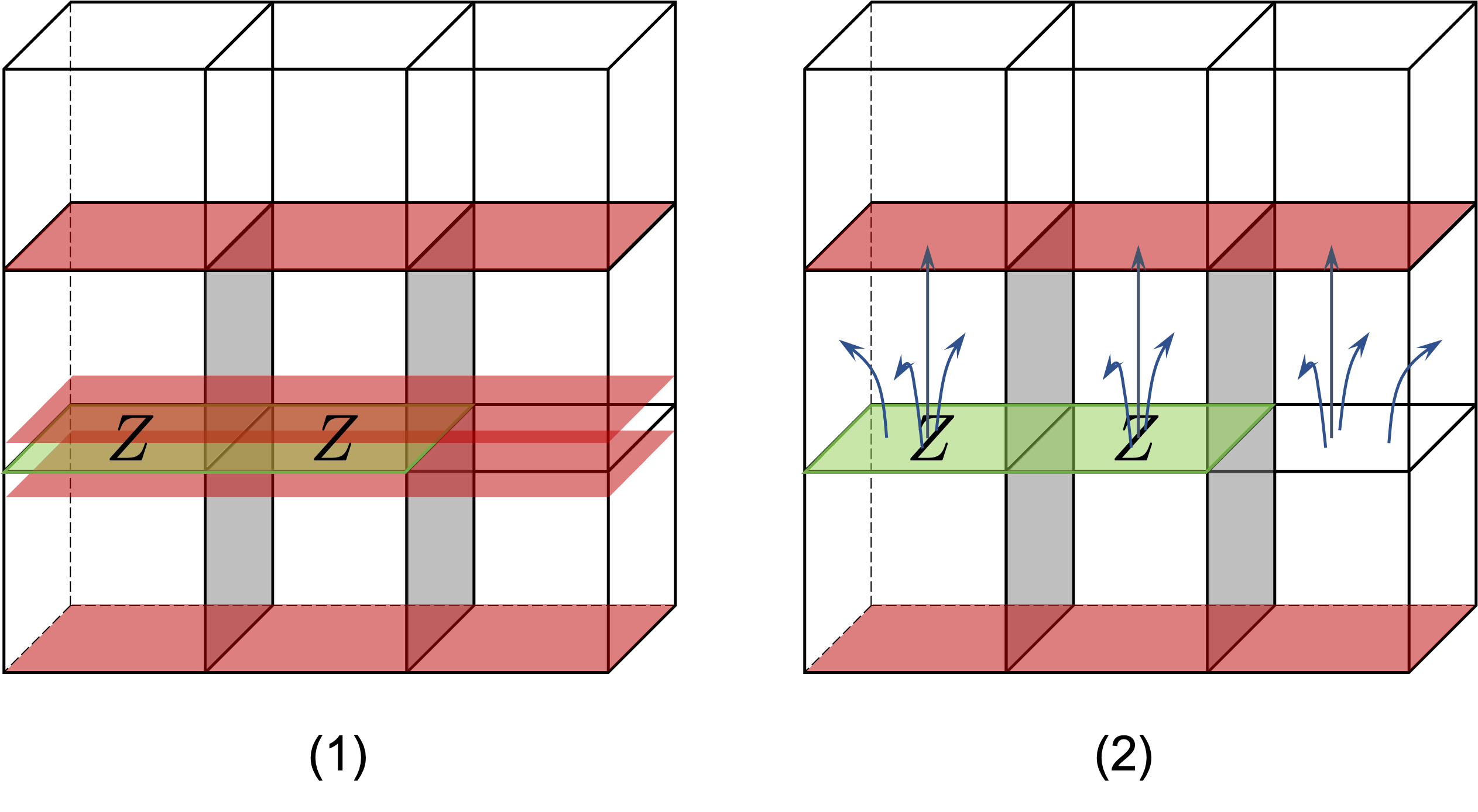}
    \caption{Fusion of two Cheshire strings with a finite depth circuit in $3+1$D. (1) Before fusion, $\prod X$ around each red rectangular cuboid measures the total charge on each string. The front and back faces of the cuboid are not shaded red for clarity. Qubits on the domain wall are coupled with $ZZ$ terms (the green plaquette pair). (2) Applying the controlled-NOT gates indicated by the arrows, the domain wall qubits are decoupled from the bulk. The $ZZ$ coupling remains, giving rise to the two fold degeneracy.}
    \label{fig:3DSe_fuse}
\end{figure}

Once a Cheshire string is created, it can be deformed ((1) to (2) or (2) to (3)) and moved ((1) to (3)) with finite depth circuits. Fusion of two Cheshire strings proceeds in a similar way as in $2+1$D. As shown in Fig.~\ref{fig:3DSe_fuse} (1), when two Cheshire strings (gray shaded plaquettes) lie next to each other, the plaquettes on the domain wall between them couple with $ZZ$ terms. The operator $\prod X$ around each red rectangular cuboid measures the total charge on each string. (The front and back faces of the cuboid are not shaded red for clarity.) Applying the controlled-NOT gates indicated by the arrows in Fig.~\ref{fig:3DSe_fuse}(2), the domain wall qubits are decoupled from the topological bulk. The $ZZ$ coupling remains, giving rise to a two fold degeneracy of $|00...0\rangle$ and $|11...1\rangle$, each of which merges with the two condensates into one condensate. 

\section{summary}
\label{sec:summary}

By discussing the generation and fusion of the Cheshire string using unitary circuits, we put it under the same framework as other excitations of a topological system. Table~\ref{table:excitations} summarizes the different cases. We expect that higher dimensional descendant excitations can also be systematically generated by sequential circuits with linear depth in their corresponding dimensions.

% Table

\begin{table}[t]
\centering
\begin{tabular}{ | c || c | c |}
\hline
topo. excitations & generated by   &  equivalence \\ \hline \hline
abelian anyon & FD 1D & LU \\ \hline
nonabelian anyon & SLD 1D &  LU \\ \hline
abelian flux loop & FD 2D & 
 FD 1D \\  \hline
non-abelian flux loop & SLD 2D & FD 1D \\ \hline \hline
\multirow{2}{*}{invertible 1D descendant} & FD 2D or & \multirow{2}{*}{FD 1D} \\ 
 & SLD 1D  & \\ \hline
noninvert. 1D descendant & SLD 1D  & FD 1D \\\hline
\end{tabular}
 \caption{Generation and equivalence of excitations in topological states with unitaries. LU refers to local unitary; FD refers to finite depth circuit; SLD refers to sequential linear depth circuit. The upper section is for elementary excitations and the lower section is for descendant excitations.}
 \label{table:excitations}
\end{table}

Anyons in $2+1$D and gauge charge/gauge flux excitations in $3+1$D are generated as the boundary of a higher dimensional unitary operator. Anyons and gauge charges are generated as the end point of a string operator while gauge fluxes are generated as the boundary of a membrane operator. When the excitation is abelian, the string/membrane operator can be implemented with a finite depth circuit. In other words, small pieces of the string/membrane operator can be connected without defect. When the excitation is non-abelian, the string/membrane operator has to be implemented sequentially, and requires a sequential linear depth circuit in 1d/ 2d. Such excitations are called elementary excitations. 

Descendant excitations like Cheshire string on the other hand do not have to be created as a closed loop and hence can live on open strings (or higher dimensional discs). They are nontrivial in the sence that when they are created in the dimension they are in, a sequential linear depth circuit is needed while trivial excitations are created with finite depth circuits. Among the descendant excitations, some are invertible, such as SPT states and Majorana chains, etc. The invertible excitations can also be created as the boundary of a higher dimensional unitary and in this case a finite depth circuit is enough. The non-invertible ones, on the other hand, cannot be created with finite depth circuit even as the boundary of one higher dimension. 

Once the excitations are created, their equivalence is established by local unitary operations if the excitation is 0d and by finite depth circuits of $n$d if the excitation is $n$d. That is, the excitations can be deformed, moved, and fused using such unitaries. 

\begin{acknowledgments}
We are indebted to inspiring discussions with Liang Kong, John McGreevy and Xiao-Gang Wen. X.C. is supported by the National Science Foundation under award number DMR-1654340, the Simons collaboration on ``Ultra-Quantum Matter'' (grant number 651438), the Simons Investigator Award (award ID 828078) and the Institute for Quantum Information and Matter at Caltech. N.T. and X.C. are supported by the Walter Burke Institute for Theoretical Physics at Caltech. 
\end{acknowledgments}
\newpage

\bibliography{references}

\appendix

\section{Other types of defects in 2+1D Toric Code}
\label{sec:defects}

Following the notation in Ref.~\onlinecite{roumpedakis2022higher}, the Cheshire string discussed in section~\ref{sec:2Dgenerate} and \ref{sec:2Dfuse} is a defect of the $2+1$D Toric Code labeled by $S_e$. There are four more types of nontrivial defects in $2+1$D Toric Code -- $S_m$, $S_{\psi}$, $S_{em}$ and $S_{me}$. Their generation and fusion follow very similar rules. To generate these nontrivial descendant defects on an open interval, a linear depth sequential circuit is needed. To deform, move or fuse these defect, a finite depth circuit is sufficient. In this section, we demonstrate explicitly the generation of $S_m$, $S_{\psi}$ and the fusion of $S_{em}\times S_e = S_{e}$, $S_{\psi}\times S_e = S_m$ and $S_{\psi}\times S_{\psi} = S_{1}$, where $S_{1}$ is the trivial defect. All other generation and fusion processes can be derived from here. For the fusion process, we will show how the circuit works in the bulk of the defect without involving the end points. The end points usually lead to extra complications but do not change the fusion result.

In all figures, dashed black edges are in the $|0\rangle$ state stabilized by $Z$ and dash-dotted black edges are in the $|+\rangle$ state stabilized by $X$. No-arrow connectors represent the controlled-$Z$ gate:
\begin{equation}
    U_{CZ} = \begin{pmatrix} 1 & 0 & 0 & 0 \\ 0 & 1 & 0 & 0 \\ 0 & 0 & 1 & 0 \\ 0 & 0 & 0 & -1
    \end{pmatrix}.
\end{equation}
One-arrow connectors represent the controlled-$X$ gate with the arrow pointing to the target:
\begin{equation}
    U_{CX} = \begin{pmatrix} 1 & 0 & 0 & 0 \\ 0 & 1 & 0 & 0 \\ 0 & 0 & 0 & 1 \\ 0 & 0 & 1 & 0
    \end{pmatrix}.
\end{equation}
Two-arrow connectors represent the $X$-controlled-Not gate, which is the controlled-$X$ gate with the control qubit conjugated by the Hadamard gate:
\begin{equation}
    U_{XCX} = \frac{1}{2}\begin{pmatrix} 1 & 1 & 1 & -1 \\ 1 & 1 & -1 & 1 \\ 1 & -1 & 1 & 1 \\ -1 & 1 & 1 & 1
    \end{pmatrix}.
    \label{eq:XCX}
\end{equation}

\begin{figure}[t]
    \centering
    \includegraphics[scale=0.43]{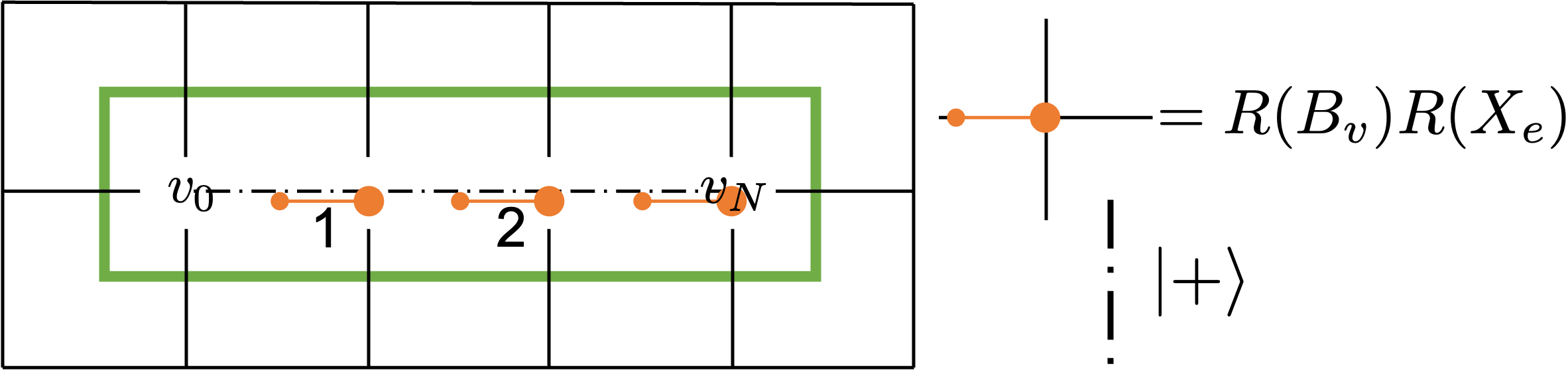}
    \caption{Generation of $S_m$ with a sequential linear depth circuit. The orange gate sets are applied sequentially from left to right. $\prod Z$ around the green box measures the total flux on $S_m$.}
    \label{fig:Sm}
\end{figure}

The gauge flux $m$ is dual to the gauge charge $e$ in Toric Code by switching between lattice and dual lattice and between $X$ and $Z$ operators. Therefore, to generate the $S_m$ defect from vertex $v_0$ to vertex $v_N$, a sequential circuit of the form
\begin{equation}
U = \prod_{i=N}^1 R(B_{v_i})R(X_{e_{i-1,i}})
\end{equation}
as shown in Fig.~\ref{fig:Sm} can be applied, where again $R(O) = e^{-i\frac{\pi}{4}O}$. This circuit maps $B_{v_i}$ to $X_{e_{i-1,i}}$ while leaving all plaquette Hamiltonian terms invariant. It hence opens up a slit of vacuum state with a `rough' boundary to the topological bulk. 

\begin{figure}[b]
    \centering
    \includegraphics[scale=0.50]{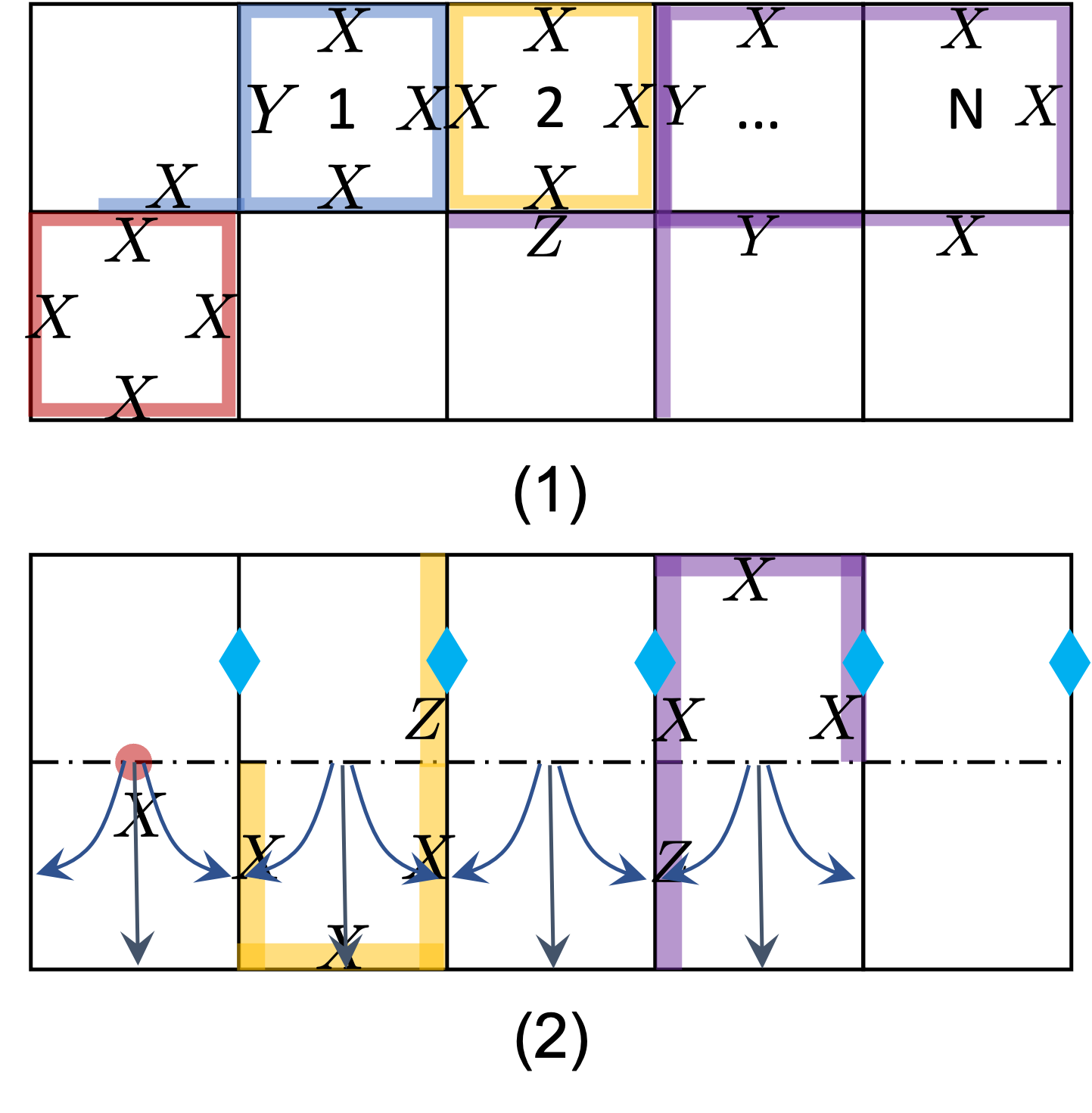}
    \caption{Generation of $S_{\psi}$ with a sequential linear depth circuit. The five-body blue term is used for the sequential part of the circuit. Following that there are two finite depth steps: the controlled-$X$ gates represented by the arrows and the Hadamard gate represented by the blue diamonds. Hamiltonian terms before and after the circuit are shown with corresponding colors (red, yellow and purple). }
    \label{fig:Ssi}
\end{figure}

The $S_{\psi}$ defect is an invertible domain wall inside the topological bulk which permutes $e$ and $m$ excitations. That is, if an $e$ excitation goes through the defect, it comes out as $m$ and vice versa. The $S_{\psi}$ defect can be generated with a sequential circuit as shown in Fig.~\ref{fig:Ssi}. Denote the blue term in Fig.~\ref{fig:Ssi}(1) around plaquette $1$ as $O_1$. The first step of the circuit is
\begin{equation}
U = \prod_{i=N}^1 R(O_i)
\end{equation}
This step is a sequential linear circuit. The second step is composed of controlled-X gates represented by the one-arrow connectors in the bottom plaquettes, as well as $R(Z)$s represented by blue diamonds, as shown in Fig.~\ref{fig:Ssi}(2). The gates in the second step all commute with each other, therefore the second step has depth one. Hamiltonian terms before and after the circuit are shown with corresponding colors (red, yellow and purple) in (1) and (2). The resulting terms take the same form as in Fig. 1 of Ref.~\onlinecite{Kitaev2012}. A different version of the circuit was proposed in Ref.~\onlinecite{Lensky2023} and implemented in Ref.~\onlinecite{Google2023}.

\begin{figure}[t]
    \centering
    \includegraphics[scale=0.37]{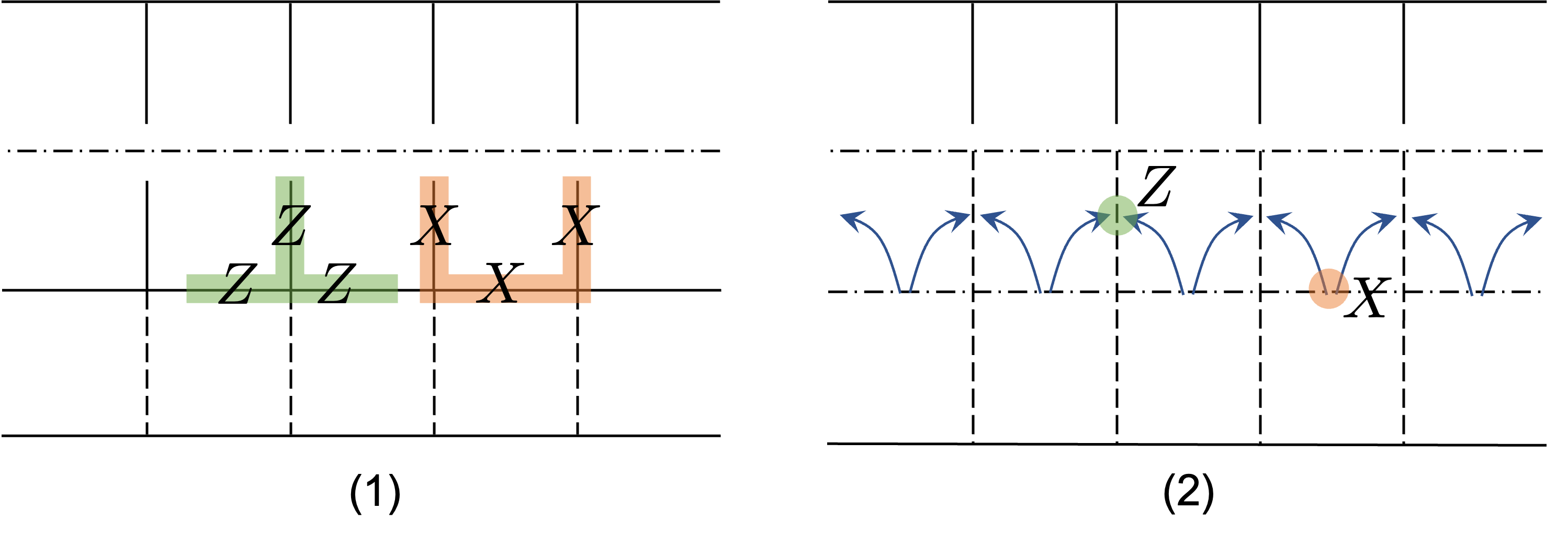}
    \caption{Fusion of $S_{e}$ (bottom in (1)) with $S_m$ (top in (1)) into $S_{em}$ ((2)) with a finite depth circuit. The circuit is composed of commuting controlled-Not gates represented by the arrows. All qubits on the domain wall between the two defects are decoupled after the circuit.}
    \label{fig:SeSm}
\end{figure}

Fig.~\ref{fig:SeSm} shows the fusion of $S_e$ and $S_{m}$ into $S_{em}$. In Fig.~\ref{fig:SeSm} (1), the bottom defect is the $S_e$ defect with smooth boundaries on the two sides and the top defect is the $S_m$ defect with rough boundaries on the two sides. Applying the controlled-Not gates represented by the one-arrow connectors in (2) maps the three body Hamiltonian terms on the domain wall between the two defects (shown in (1)) to single $X$ and $Z$ terms (shown with corresponding color in (2)). Therefore, after the finite depth circuit, the two defects merge into one, $S_{em}$, with smooth boundary at the bottom and rough boundary at the top.

\begin{figure}[b]
    \centering
    \includegraphics[scale=0.38]{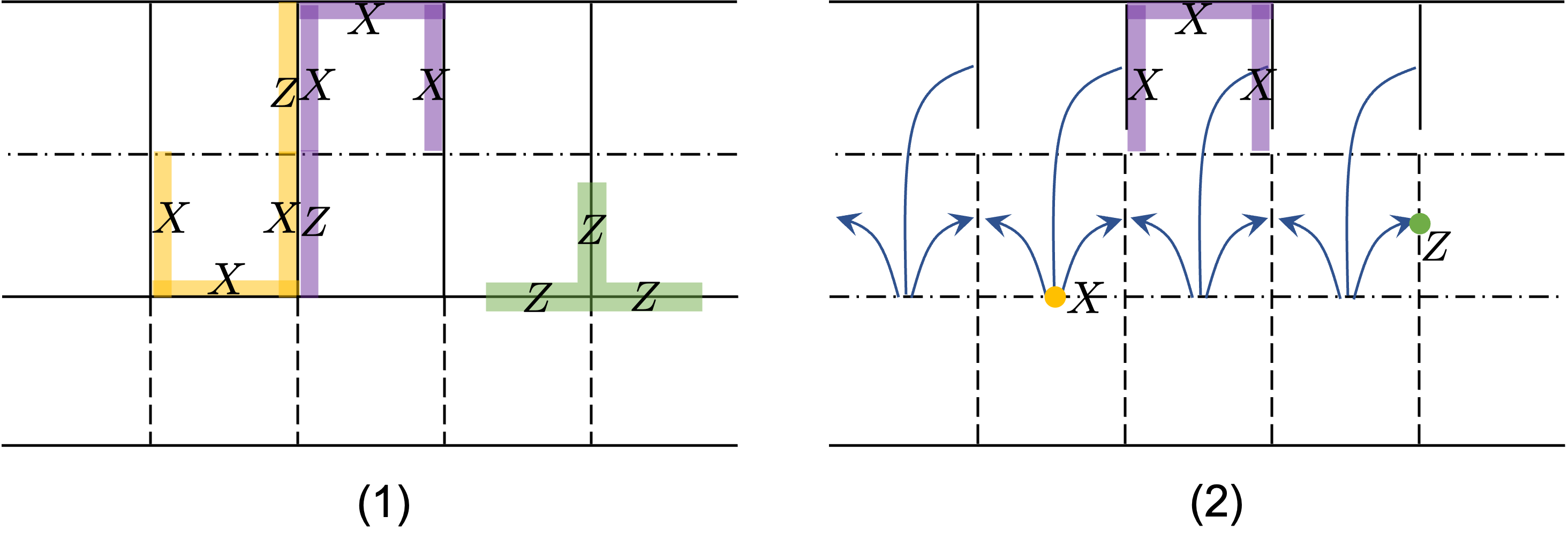}
    \caption{Fusion of $S_e$ (bottom in (1)) with $S_{\psi}$ (top in (1)) into $S_{em}$ ((2)) with a finite depth circuit. The circuit is composed of commuting controlled-$Z$ gates (represented by the no-arrow connectors) and controlled-Not gates (represented by the one-arrow connectors.) Yellow and green terms map to decoupled qubits while the purple terms map to half plaquette terms on the rough boundary.}
    \label{fig:SeSsi}
\end{figure}

Fig.~\ref{fig:SeSsi} shows the fusion of $S_e$ and $S_{\psi}$ into $S_{em}$. In Fig.~\ref{fig:SeSsi}(1), the bottom defect is the $S_e$ defect with smooth boundaries on the two sides and the top defect is the $S_{\psi}$ defect taking the form shown in Fig.~\ref{fig:Ssi}. The circuit is composed of controlled-$Z$ gates (represented by the no-arrow connects) and the controlled-Not gates (represented by the one-arrow connectors). All gates commute and the circuit has depth one. The yellow and green terms are mapped to single qubit terms after the circuit while the purple term becomes a three body plaquette term on the rough side of the boundary. Therefore, $S_e$ and $S_{\psi}$ fuse into $S_{em}$, as shown in Fig.~\ref{fig:SeSsi}(2), which take the same form as in Fig.~\ref{fig:SeSm}(2)

\begin{figure}[t]
    \centering
    \includegraphics[scale=0.34]{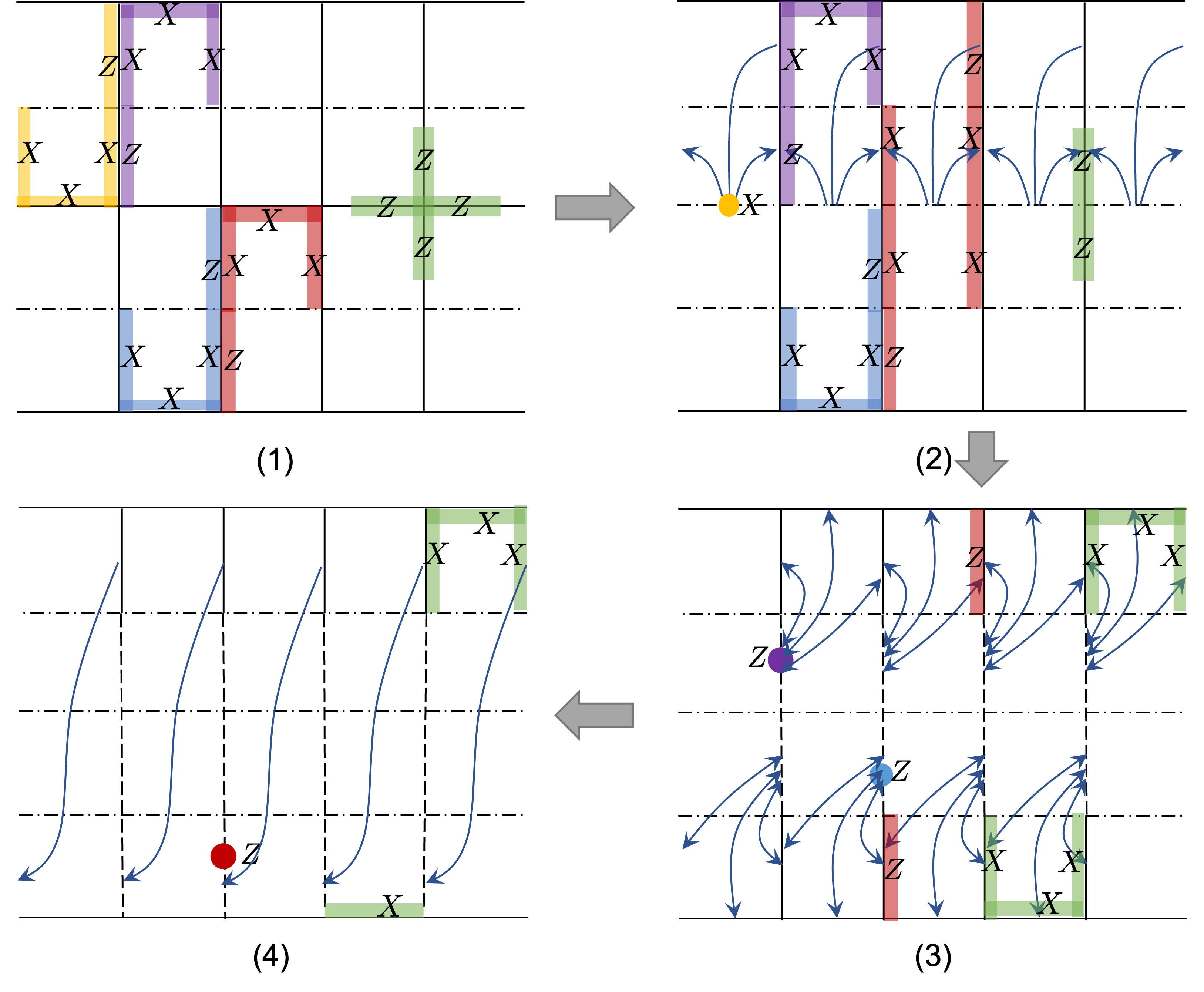}
    \caption{Fusion of $S_{\psi}$ with $S_{\psi}$ into $S_{1}$ with a finite depth circuit. The circuit is composed of controlled-$Z$ gates (the no-arrow connectors), controlled-Not gates (the one-arrow connectors) and the $X$-controlled-$X$ gates (the two-arrow connectors). In (4), all qubits in the middle are decoupled while qubits on the two sides connect to form complete plaquette terms of the Toric Code.}
    \label{fig:SsiSsi}
\end{figure}

$S_{\psi}$ is an invertible defect and $S_{\psi}\times S_{\psi} = S_1$. Fig.~\ref{fig:SsiSsi} shows how the fusion can be realized with a finite depth circuit. The first step of the circuit ((1) to (2)) is composed of controlled-$Z$ gates (represented by the no-arrow connects) and the controlled-Not gates (represented by the one-arrow connectors) as shown in (2). After this step the yellow terms map to decoupled qubits in $|+\rangle$ state on the middle line. The second step of the circuit ((2) to (3)) is composed of $X$-controlled-Not gates (Eq.~\ref{eq:XCX}) represented by the two-arrow connectors in (3). The purple and blue terms maps to decoupled qubits in this step. Finally, with controlled-Not gates in the last step (shown in (4) with the one-arrow connectors), the red terms map to decoupled qubits. The green terms become the plaquette term of the Toric Code bulk. All gates in each step of the circuit commute with each other, therefore the circuit has finite depth.

\begin{figure}[ht]
    \centering
    \includegraphics[scale=0.35]{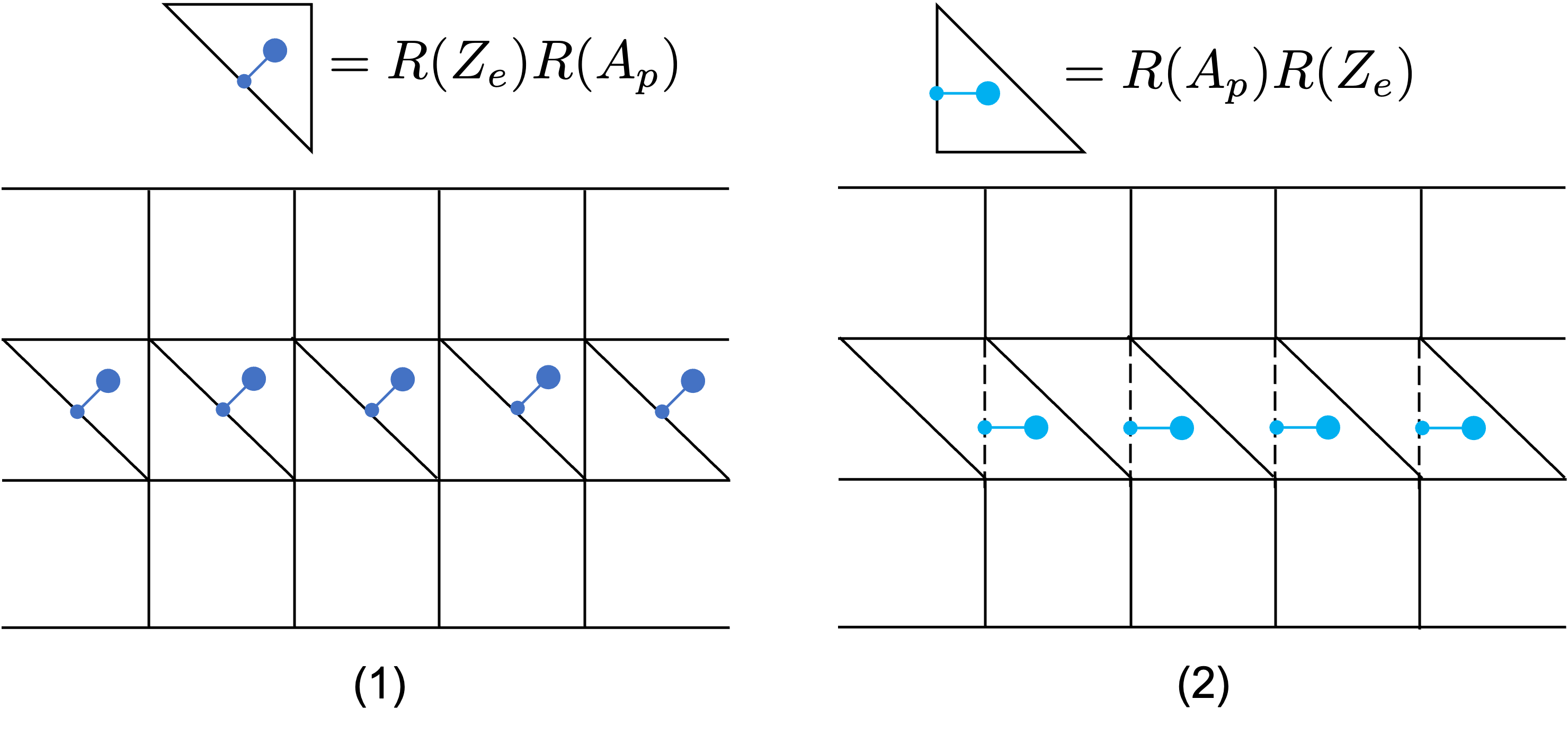}
    \caption{Creating a dislocation in Toric Code with a finite depth circuit. In step 1, the dark blue gate sets are used to add diagonal edges to divide square plaquettes into triangles. In step 2, the light blue gate sets are used to remove vertical edges and merge two triangles into a parallelogram. A dislocation line is generated after these two steps.}
    \label{fig:dislocation}
\end{figure}

It might seem that instead of recovering the regular Toric Code on square lattice, we end up with a dislocation on the square lattice. But this is not a problem because a dislocation can be generated or removed with finite depth circuit in Toric Code as shown in Fig.~\ref{fig:dislocation}. 

Starting from a regular Toric Code on square lattice, diagonal edges can be added into each plaquette with the dark blue gate sets as shown in Fig.~\ref{fig:dislocation} (1), dividing each square into two triangles. All the dark blue gate sets commute with each other and can be applied in one step. Next, the vertical edges between the triangles can be removed with the light blue gate sets as shown in Fig.~\ref{fig:dislocation} (2), merging two triangles into a parallelogram. All light blue gate sets commute with each other as well, so we have another depth one circuit. After these two steps, we have introduced a dislocation defect into the square lattice.

\end{document}